\documentclass[12pt]{article}
\usepackage{epsfig}
\let\Large=\large
\let\large=\normalsize

\newcommand{\be}[3]{\begin{equation}  \label{#1#2#3}}     
\newcommand{\ee}{ \end{equation}}
\newcommand{\ba}{\begin{array}}
\newcommand{\ea}{\end{array}}
\newcommand{\NP}[3]{{\em Nucl. Phys.}{ \bf B#1#2#3}}

\renewcommand{\arraystretch}{1.8}

\begin{document}

\thispagestyle{empty}
\rightline{HUB-EP-97/40}
\rightline{SU-ITP-97-35}
\rightline{hep-th/9708065}

\vspace{1truecm}
\centerline{\bf \Large Moving Moduli, Calabi-Yau Phase Transitions and} 
\vskip0.3cm
\centerline{\bf \Large
Massless BPS Configurations in Type II Superstrings}
\vspace{1.2truecm}
\centerline{\bf Klaus Behrndt\footnote{e-mail: 
 behrndt@qft2.physik.hu-berlin.de}}
\vspace{.5truecm}
\centerline{\em Department of Physics, Stanford University}
\centerline{\em Stanford, CA 94305, USA}

\vspace{1truecm}

\centerline{\bf Dieter L\"ust\footnote{e-mail: 
 luest@qft1.physik.hu-berlin.de} and Wafic A. Sabra\footnote{e-mail:
 sabra@qft2.physik.hu-berlin.de}}
\vspace{.5truecm}
\centerline{\em Humboldt-Universit\"at, Institut f\"ur Physik}
\centerline{\em Invalidenstra\ss e 110, 10115 Berlin}
\centerline{\em Germany}

\vspace{1.2truecm}


\vspace{.5truecm}

\begin{abstract}
In this paper we discuss compactifications of type II superstrings where
the moduli of the internal Calabi-Yau space vary over four-dimensional
space time. The corresponding solutions of four-dimensional $N=2$
supergravity are given by charged, extremal  BPS black hole
configurations with non-constant scalar field values. In particular we
investigate the behaviour  of  our solutions near those points in
the Calabi-Yau moduli space where some internal cycles collapse and
topology change (flop transitions, conifold transitions) can take place.
 The singular loci in the internal space are related to special points
in the uncompactified space. The phase transition can happen either at
spatial infinity (for positive charges) or on spheres (with at least one
negative charge). The corresponding BPS configuration has zero ADM
mass and can be regarded as a domain wall that separates
topologically different vacua of the theory.

\end{abstract}

\bigskip \bigskip
\newpage

\noindent
{\bf \large 1. Introduction} 

\bigskip 

Recent developments showed that space-time geometry or space-time
topology have no absolute meaning in string theory but are derived
concepts which depend on how strings test their background spaces in
various limits. For example, in the study of the moduli spaces of type II
strings compactified on Calabi-Yau three-folds it became clear that the
topologies of the Calabi-Yau spaces can be continuously deformed into
each other; therefore many, or even possibly all, Calabi-Yau vacua are
just branches of a larger universal moduli space.  Calabi-Yau
transitions, which were studied in the past, contain for example flop
transitions where one moves through walls of the complexified K\"ahler
moduli spaces (see for example \cite{aspgremor}, \cite{witten2}) where
the walls correspond to metrically degenerate Calabi-Yau spaces in
which certain homologically nontrivial cycles have zero
volume. Performing the flop transition the Calabi-Yau intersection
numbers change, while the number of K\"ahler moduli is left unchanged.
However the conformal field theory \cite{witten} stays perfectly well
behaved when moving through a wall. Another transition is the conifold
transition in the complex structure moduli space of type IIB
superstrings on Calabi-Yau spaces. The conifold transitions
\cite{conifolds} occur at those points in the moduli space where
certain 3-cycles shrink to zero size and then are blown up as two
cycles, changing in this way the Hodge numbers.  However, unlike the
previous case, the conifold transition cannot be described in the
language of conformal field theory.  The physical understanding of the
conifold transition was provided by Strominger \cite{strominger}; at
the conifold point a BPS hypermultiplet black hole becomes massless,
being responsible for the singularity in the moduli space metric of
the $N=2$ vector multiplets at this point. The massless black can
condense \cite{gremorstro}, i.e. acquire a non-vanishing vacuum
expectation value (just in analogy to the monopole condensation in
$N=2$ field theory \cite{sw}), and in this way one branches off into
the moduli space of a Calabi-Yau space with different Hodge numbers
($h^{(1,1)}$ gets larger, whereas $h^{(2,1)}$ gets reduced).

BPS saturated solutions of four-dimensional $N$=2 supergravity 
coupled to $N=2$ vector multiplets have been discussed in
many recent papers \cite{be/ca,be/mo,sa,be/lu}. 
The simplest class of solutions is given by the double extreme $N=2$
black holes with non-vanishing electric and magnetic charges.
For this type of solutions the values of the scalar moduli fields, which 
follow from a minimisation of the $N=2$ central charge, 
take constant values
over the entire space-time. More general extremal supersymmetric
black holes allow for non-constant moduli fields, where the variation
of the moduli fields over the four-dimensional space is provided by
those harmonic functions which also determine the electric/magnetic
field strengths of the charged black hole \cite{sa,be/lu}. So in this case
of non-constant moduli, we have learned that the internal space
does not decouple from the 4-dimensional space time.  
In particular the vector multiplet moduli, which determine 
the K\"aher moduli in type IIA Calabi-Yau compactifications
or which, respectively, determine the sizes
of the Calabi-Yau 3-cycles in the type IIB compactifications, vary
over the uncompactified space in case of static extremal $N=2$
black hole solutions. Indeed
one can argue that special or singular points in the internal space are
related to special or singular points in space time (like horizons or
curvature singularities). It is the aim of this paper to discuss this
strong relationship.

After discussing the general $N=2$ supergravity solution, we will
focus on the interplay between space time and internal space, which is
assumed to be a Calabi-Yau-threefold of a type II compactification.
First we will briefly study the variation of the type IIA K\"ahler
moduli in the neighbourhood of the boundary of the complexified
K\"ahler cone, i.e. for small value of a particular K\"ahler
modulus. This discussion is relevant for the flop transition in
four dimensions. Then we will more extensively discuss the type IIB conifold
transition where one of the complex structure moduli fields is small.
The following generic picture will emerge. If one moves through a non-singular
space time one varies at the same time the radii of the cycles of the
Calabi-Yau. At any point in space time the Calabi-Yau looks
different.  Any generic $N$=2 black hole has a non-singular horizon
and this horizon in space time corresponds to an extremal radius of
the Calabi-Yau. We will find that for vanishing 3-cycles in IIB
compactifications, our
solution contains massless BPS states.

\bigskip

{\bf \large 2. The self-dual 3-brane}

Before we come to the Calabi-Yau compactification, let us explain
the picture for the torus compactification. 

The self-dual 3-brane of type IIB string theory in 10 dimensions
is given by
\be210
\ba{c}
ds^2 = {1 \over \sqrt{H}} (-dt^2 + dz_1^2 + dz_2^2 + dz_3^2) + 
\sqrt{H} (dy_1^2 + dy_2^2 + dy_3^2 + dx_i^2) \ , \\
F = d {1 \over H} \wedge dt \wedge dz_i + {^{\star}d} 
{1 \over H} \wedge dt \wedge dz_i \ 
\ea
\ee
where $i=1, 2,3$ with the harmonic function $H$
\be220
H = 1 + {p \over r} \qquad , \qquad r^2 = x_i x_i \ ,
\ee
and $p$ is the charge.  In general, the harmonic function can depend
on all transversal coordinates $(y_i,x_i)$ yielding a non-singular
space time.  But since we are interested in a compactification to 4
dimensions, we assumed that $H$ is independent of all internal
coordinates $(y_i,z_i)$, i.e.\ not only of world volume coordinates
($z_i$) but also of the transversal coordinates ($y_i$).  This
assumption makes already the 10-d solution singular. Let us look in
more detail on this solution. Compactifying this solution means we
wrap the 3-brane coordinates ($z_i$) around one 3-cycle and the
remaining coordinates ($y_i$) represent a second 3-cycle.
Both 3-cycles build the internal space, and their radii vary over space
time. By approaching the singular point $r=0$ one 3-cycle diverges
whereas the other vanishes. 


The singularity in this case is a consequence of missing charges.
In torus compactification we need at least 4 charges
(or 4 branes) in order to obtain a non-singular black hole in
4 dimensions. The corresponding solution with four intersecting 3-branes
is given by (we skip the gauge field part)
\be230
\ba{rcl}
ds^2 &=& -{1 \over \sqrt{H_1 H_2 H_3 H_4}} dt^2 +  \sqrt{H_1 H_2 H_3 H_4}
dx_i dx_i + \\&&  + 
\sqrt{H_1 H_2 \over H_3 H_4} dz_1^2 + \sqrt{H_3 H_4 \over H_1 H_2}
dy_1^2 + \sqrt{H_1 H_4 \over H_3 H_2}(dz_2^2 + dz_3^2) + 
\sqrt{H_3 H_2 \over H_1 H_4} (dy_2^2 + dy_3^2) \ .
\ea
\ee
($H_m = 1 + p^m/r,\ m=1,\cdots,4$) and the branes are wrapped around the 
following internal coordinates 
\renewcommand{\arraystretch}{1}
\be240
\label{070}
\ba[c]{lcccccccc}
\hline
 &  z_1 \ & \ z_2\  & \ z_3\  & \ y_1 \ & \ y_2 \ & \ y_3  \\
\hline
H_1 - 3-brane \qquad & & & & \times & \times &  \times \\
H_2 - 3-brane &  & \times & \times & \times &  \\
H_3 - 3-brane & \times & \times & \times &  &  & \\
H_4 - 3-brane & \times &  &  & & \times &\times  \\
\hline
\ea
\ee 
where the world-volume coordinates are indicated by ``$\times$''.
Now, all four 3-cycles stay finite when we approach the
point $r=0$ and also the 4-d geometry is not singular at this
point ($AdS_2 \times S_2$). But turning off one charge, 
e.g.\ $p^1$ (i.e.\ $H_1=1$), the configuration becomes singular, some
3-cycles vanish whereas others diverge yielding a
decompactification at $r=0$. This is typical for torus
compactifications. It is not possible to shrink one cycle while keeping
the others fixed.
\renewcommand{\arraystretch}{1.8}

The ADM mass for the solution (\ref{230}) is
\be250
M = {1 \over 4} (p^1 + p^2 + p^3 + p^4) \ .
\ee
Hence, for a vanishing mass we need at least one negative charge.
However, this yields a zero in the corresponding harmonic function ($H=
1- {|p|\over r}$) at $r=|p|$. As a consequence one 3-cycle vanishes, but
again the others diverge. 

The situation changes completely when we compactify the 3-brane
configuration on a Calabi-Yau threefold. Here, any generic solution has
stabilized radii and a non-singular horizon. 
Even for vanishing 3-cycles, the other cycles remain non-singular 
(also the horizon), but the gauge
couplings will diverge. It will be the subject of the next sections to
address this question.

\bigskip

{\bf \large 3. Extremal black holes in $N$=2 supergravity}

We start with a short review of the general black hole solution 
of $N$=2 supergravity coupled to vector multiplets.
Assuming that the scalar fields in the  hyper multiplets are 
trivial,\footnote{For solutions with non-trivial hyper multiplets see
\cite{hyper}.}
the supersymmetric solution for the
black hole metric, the scalar field components $z^A$ and 
for the magnetic/electric fields strenths $G_{I\, mn}$,
$F^I_{mn}$ is given by \cite{sa}, \cite{be/lu}
\be310
\ba{c}
ds^2 =  - e^{2U} dt + e^{-2 U} dx^m dx^m \qquad , \qquad z^A = 
{X^A \over X^0}\\
F^I_{mn} = {1 \over 2} \epsilon_{mnp} \partial_p \tilde{H}^I
\qquad , \qquad G_{I\, mn} = {1 \over 2} \epsilon_{mnp} \partial_p H_I \ ,
\ea
\ee
with $m,n=1,2,3$ and
\be320
\ba{rcl}
G_{I\, \mu\nu} & = &\mbox{Re} {\cal N}_{IJ} F^{J}_{\mu\nu} - 
 \mbox{Im} {\cal N}_{IJ} {^{\star}}F^J_{\mu\nu} \ , \\
e^{-2U}=e^{-K}  & \equiv & i (\bar{X}^I F_I- X^I \bar{F}_I) \ , 
\ea
\ee
where $K$ is the K\"ahler potential of the moduli space metric, and
the holomorphic section $(X^I , F_I)$ are constrained by
\be330
i (X^I - \bar{X}^I ) = \tilde{H}^I(x^{\mu}) \qquad , \qquad 
i (F_I -\bar{F}_I) = H_I(x^{\mu})  \ .
\ee
Since the symplectic vector $( \tilde H^I , H_I )$ is introduced
via the gauge fields, this condition can be seen as a relation 
between the holomorphic section $(X^I , F_I)$ and the gauge field
section $( F^I_{\mu\nu}, G_{\mu\nu\, I})$. As a consequence of the 
Bianchi identities, the $H's$ have to be
harmonic and for the single-center case we take
\be340
\tilde H^I = \tilde h^I +{ p^I \over r} \qquad , \qquad 
H_I =  h_I + { q_I \over r} 
\ee
where $q_I$ and $p^I$ are the electric and magnetic charges
carried by the black hole. The constant parts determine the
scalar fields at infinity and parameterize the moduli space.

In addition, we have the two constraints
\be350
( H_I \partial_m \tilde{H}^I - \tilde{H}^I \partial_m H_I) =0 \quad , \quad
e^{-2U_{\infty}} = 1 \ .
\ee
The first condition ensures that the K\"ahler connection
vanishes and is necessary for any static solution (see \cite{sa},\cite{be/lu})
and the second one fixes our coordinate system. 
These two conditions fix two of the $h's$ and we have
the total number of $2n$ continuous variable parameterizing
the $2n$ dimensional moduli space. 

In order to discuss the singularities of the black hole metric  (\ref{310})
we compute the square of the space-time curvature tensor
\be510
\ba{c}
R_{\mu\nu\rho\sigma}\,R^{\mu\nu\rho\sigma} =
{1\over V^8} \left( 32 \, (\partial_m V \partial_m V)^2 - 48 \, V
\partial_n \partial_m V  \, \partial_m V  \, \partial_n V  - 2 \,
V\, \partial_m V \partial_m V  \, \partial_n \partial_n V + \right.\\ 
\left. + 8 \, V^2 \, \partial_n \partial_m V \, \partial_n\partial_m V  
+ V^2 (\partial_m\partial_m V )^2 
\right),
\ea
\ee 
where $V=e^{-U}= e^{-K/2}$.

\bigskip

{\bf \large 4. The solution near a wall in the Calabi-Yau 
K\"ahler moduli space}

In this section we want to construct the black hole solutions in the
limit where one of the $h^{(1,1)}$ K\"ahler moduli 
$z^A$ ($A=1,\dots ,h^{(1,1)}$) in type IIA compactifications
on a Calabi-Yau space becomes small. 
These moduli correspond to sizes of 2-cycles, respectively of the
Hodge dual 4-cycles in the Calabi-Yau threefold; hence they must be
positive defining the K\"ahler cone. If one of the K\"ahler moduli
becomes small  the Calabi-Yau space degenerates in one of the
following three ways \cite{witten2}

\noindent (i) A two cycle collapses to a point.

\noindent (ii) A complex divisor collapses either to  a curve or to a point.

Case (i) corresponds to a topology change via a flop transition
between two different but birationally equivalent
Calabi-Yau spaces, where 
by moving through the wall a new geometrical
K\"ahler cone is reached; the union of all geometrical K\"ahler cones
is called the extended K\"ahler cone. During the flop transition
one K\"ahler modulus, say $z^2$ changes its sign and the size of the new
2-cycle is given by $-z^2$. This has the effect that the intersection
numbers change by the new term $C_{222}=-{1\over 6}$ \cite{aft}.
(Several concrete examples of flop transitions in particular Calabi-Yau
spaces where investigated in \cite{louis}.)
However in four-dimensional type IIA compactifications on a Calabi-Yau
space, the flop transition is in fact not a sharp transition since one
can turn on the axionic components of the complex moduli fields. We will
discuss in the following the case of vanishing axions.
Moreover in four-dimensions there are in general non-geometrical phases outside
the extended K\"ahler cone due to the effect of world sheet
instantons. In the following we will neglect all effects coming from
world sheet instantons. This resembles in a way the decompactification
to five dimensions or $M$-theory on a Calabi-Yau manifold \cite{fkm},
since in five dimensions the axions fields are frozen and no non-geometrical
phases exist \cite{witten2} due to the absence of world sheet instantons.
Phase transitions via flops in 5-dimensional compactifications
with constant moduli fields were recently discussed in \cite{chou}.

Now let us discuss the extremal $N=2$ black holes for small values
of the K\"ahler moduli. 
At the vicinity of the K\"ahler cone  wall, i.e. for small K\"ahler
moduli, we expand the $N=2$ prepotential as 
\be999
F(X^I)=(X^0)^2\lbrack -{1\over 6}C_{ABC} z^A z^B z^C-{i\chi\zeta 
(3)\over 2(2\pi )^3}\rbrack \ ,
\ee
where the $z^A=X^A/X^0$ are the K\"ahler moduli 
and the $C_{ABC}$ are the classical intersection numbers.
Assuming that $z^2$ corresponds to the vanishing
cycle, the intersection numbers change via the flop transition
\be705
- {1 \over 6} C_{ABC}z^A z^B z^C  \rightarrow -{1 \over 6} C_{ABC} 
z^A z^B z^C + {1 \over 6} (z^2)^3
\ee
Next, we have to solve the constraints (\ref{330}). An easy way to find
solutions is to restrict ourselves to the axion-free case, i.e.
\be710
\mbox{Re} z^A = 0 
\ee
and find as a solution
\be720
X^0 = {\lambda \over 2} \quad , \quad  X^A = - i\, { \tilde H^A \over 2} 
\qquad \mbox{and} \qquad z^A = - i \, {\tilde H^A \over \lambda}  \ .
\ee
So we are dealing with a black hole with electric charge $q_0$ and 
non-vanishing magnetic charges $p^A$.
The function $\lambda$ is fixed by the equation $F_0 - \bar F_0 = -i H_0$.
Expanding the solution of this cubic equation in powers
of the intersection form we find
\be730
\lambda = {H_0 \over 2c} + {2 c \over (H_0)^2} ({1 \over 6} 
C_{ABC} \tilde H^A \tilde H^B \tilde H^C) \pm ..
\ee
with $c = \chi \zeta(3)/2 (2\pi)^3$. Inserting this into $e^{-2U}$ yields
\be740
e^{-2U} = {(H_0)^2 \over 4c}  - {3 c \over H_0} ({1 \over 6} 
C_{ABC} \tilde H^A \tilde H^B \tilde H^C )  \pm .. \ \ .
\ee
Following the procedure, e.g., described in the last
reference of \cite{be/mo}, one could easily calculate further corrections
to the prepotential (\ref{999}).
For a microscopic discussion see also the first ref.\ of \cite{be/mo} .

For this axion-free solution, a vanishing cycle ($z^2=0$) corresponds to a
vanishing harmonic function ($\tilde H^2=0$), see (\ref{720}).
With $\tilde H^2=\tilde h^2+{p^2\over r}$, it follows that the
relevant cycle is vanishing at the special radius  $r=-{p^2\over \tilde h^2}$.
So the charge $p^2$ must be negative ($\tilde h^2>0$). Crossing this
critical radius the flop transition takes place and $z^2$ becomes negative.
We see that, even if the intersection form vanishes on the K\"ahler wall
$z^2=0$
the $H_0$ part regularizes the solution, i.e.
the space-time black hole metric
is regular on the wall of the K\"ahler
cone. When going through the wall, we
have to take into account the change in the intersection numbers as given in
(\ref{705}) in order to get the solution on the other side. In contrast to 
the conifold transition
discussed below, the Hodge numbers do not change in this transition.
Thus, the number of vector multiplets is the same.
In the next section we will also discuss the possibility 
where the phase transition takes place at spatial infinity.

Our solution (\ref{720}) with magnetic charges $p^A$   has a
nice brane interpretation in $M$-theory. It consists of an intersection
of three $M5$-brane, which have a common string. The remaining 4
directions are wrapped around 4-cycles of the Calabi Yau space
and the charges indicate how many times the 5-branes are wrapped.
Every of these 4-cycles is parameterized by one harmonic function.
In 5 dimensions we obtain a magnetic string and $H_0$ 
parameterizes the momentum modes traveling along this string. 
In 11 dimensions these momentum modes correspond to gravitational waves,
i.e.\ pure gravity solution. It is interesting, that this gravity part
regularizes our solution, even for vanishing intersection part. Since the
constant $c$ is essentially the Euler number, this contribution is
related to $R^4$ terms of the $M$-theory, for a recent discussion see
\cite{an/fe}. Hence, the regularization of the 4-d black hole is caused
by higher curvature terms in 11 dimensions.

\bigskip

{\bf \large 5. The solution near a Calabi-Yau conifold point}

A Calabi-Yau space has $b_3=2h^{(2,1)}+2$ topologically non-trivial  
3-cycles. One  introduces a basis of 3-cycles $\{ \gamma_I , \delta^J \}$ 
($I,J = 0\,\cdots \, h^{2,1}$) such that
\be405
{\bf \gamma} \cap {\bf \delta} = - {\bf \delta} \cap 
{\bf \gamma} = {\bf 1}  \ , \ 
{\bf \gamma} \cap {\bf \gamma} = {\bf \delta} \cap {\bf \delta} =0 \ .
\ee
In terms of the holomorphic 3-form $\Omega,$ the corresponding periods 
are 
\be406
 F_I = \int_{\gamma_A} \Omega \quad , \quad X^I = 
\int_{\delta^I} \Omega \ 
\ee
where $X^I$ can be used as projective coordinates on the 
moduli space ${\cal M}$ 
of complex structures. So the parameters for the complex 
structure moduli can be defined as $z^A=X^A/X^0$ ($A=1\,\cdots , h^{(2,1)}$).

The conifold point  is in general described by a locus of co-dimension $k$
in ${\cal M}$, where $k$ cycles $X^L$ ($L=1\, .. \, k$) vanish, while the
remaining cycles stay finite.
In the following we will discuss the most simple situation with  periods
$X^0$ and $X^1$ (together with $F_0$ and $F_1$), where $X^1$ vanishes
at the conifold point and $X^0$ remains finite
(there might be other periods, but they do not influence
the results). This captures 
all the interesting
physics and can be easily generalised to the higher dimensional case.
So at the conifold point we have
\be410
z^1 = 0 \qquad \mbox{or}: \qquad \mbox{Im}\, z^1 = \mbox{Re} \, z^1 = 0 \ .
\ee
In addition, if the 3-fold is transported
about a closed loop around the singular surface the period
$F_1$ undergoes a monodromy transformation \cite{gremorstro}
\be425
X^1 \rightarrow X^1 \quad , \quad F_1 \rightarrow 
 F_1 + X^1  \ .
\ee
Therefore, near this point
the prepotential can be expanded as
\be440
F = (X^0)^2 {\cal F} = -i \, (X^0)^2 \left( c + {1 \over 4 \pi}  (z^1)^2 
\log i z^1\,  + (\mbox{analytic terms})  \right) \ .
\ee
The constant $c$ contains, e.g., the Euler number $\chi$ of the
Calabi-Yau ($c= \chi \zeta(3) / 2 (2\pi)^3$). 
With eq.(\ref{320}) it is easy to see that the K\"ahler potential is finite
at the conifold point:
\be899
e^{-K}=4|X^0|^2 \, {\chi \zeta(3)\over 2 (2\pi )^3}.
\ee
This will mean that the space-time curvature square
eq.(\ref{510}) at the conifold point will also
be finite. In contrast, the internal moduli space metric diverges at the 
conifold point.

Next, let us look at the 
structure of the space-time solution near this point. Assuming that $X^0$ is
not zero or divergent (to keep all other cycles finite), it follows that 
at the conifold point
\be420
\mbox{Re} X^1 = 0 \qquad , \qquad 2\,\mbox{Im}X^1 = \tilde H^1 = 0 \ .
\ee
For simplification, we will consider axion-free black holes.
Thus, we assume
\be450
\mbox{Re} z^1 \equiv 0
\ee
and take as solution of (\ref{330})
\be460 
X^0 = {\lambda \over 2} \qquad , \qquad X^1 = -i {\tilde H^1 \over 2}
\ee
where the function $\lambda$ is fixed by
\be470
i( F_0 - \bar F_0) = H_0 = 2 c \lambda + {(\tilde H^1)^2 \over 4\pi \, 
\lambda} \ .
\ee
Since $\lambda$ should stay finite, otherwise
the non-vanishing cycles would decompactify, 
we take for $\lambda$ the solution
\be480
\lambda = {H_0 \over 2 c } -  \, {( \tilde H^1)^2\over 4\pi \, H_0} + 
{\cal O}((\tilde H^1)^4) \ .
\ee
Keeping only the first correction, we obtain for
the function $e^{-2U}$ in the metric (\ref{310}) 
\be490
\ba{rcl}
e^{-2U}&=& i(\bar X^I F_I - X^I \bar F_I) = {\lambda \over 2} H_0 +
 {(\tilde H^1)^2 \over 4 \pi} \left( \log {\tilde H^1 \over \lambda}
+ {1 \over 2} \right)
 \\ &=& {H_0^2 \over 4c} + { (\tilde H^1)^2 \over 4\pi}  \, 
\log {2c\, {\tilde H^1}\over H_0} \pm .. \ . 
\ea
\ee
which is non-singular/non-vanishing even at points where the
3-cycle vanish (${\tilde H^1}=0$). Already the $X^0$ part in the
prepotential regularises the metric. Taking into account further
analytic terms from the expansion (\ref{440}) will not alter this
qualitative picture. But note, because of the logarithm we cannot
extend the solution beyond this point, i.e.\ formally to negative
${\tilde H^1}$. Instead, ${\tilde H^1}$ remain zero and new functions 
will appear that parameterise the new emerging cycles.

Again keeping only the first correction, our periods at infinity are 
given by
\be495
\ba{l}
X^0_{\infty} = {h_0 \over 4c} - {(\tilde h^1)^2 
\over 8 \pi\,  h_0} \quad , \quad
X^1_{\infty} = -i \, {\tilde h^1 \over 2} \\
F_0^{\infty} = -i ({h_0 \over 2}  + {c (\tilde h^1)^2 \over 8 \pi \, h_0})
\quad , \quad F_1^{\infty} = - {{\tilde h^1} \over 4\pi} \, ( {1 \over 2} + 
\log {2c {\tilde h^1} \over h_0}).
\ea
\ee
The black hole carries electric ($q_0$) and magnetic ($p^1$) 
charges, see (\ref{330}) and (\ref{340}), and
its mass can  be read off from the metric
eq.(\ref{490}) and can be written in the usual way 
\be415
M = |q_0 X^0_{\infty} - p^1 F_1^{\infty}| \ .
\ee
Taking into account more analytic terms in the prepotential (\ref{440}) 
would correspond to additional magnetic charges. 
Note, $h_0$ does not define a modulus, it is fixed by the second
constraints in (\ref{350}); the first one is identically fulfilled for
axion-free black holes. On the other hand, $\tilde h^1$ is a modulus. 
In order to justify the  expansion of the prepotential
eq.(\ref{440}) around the conifold point, $\tilde h^1$ should be small.

In order to make the picture complete we have to discuss the
options for a vanishing harmonic function, i.e.\ to fulfill
equation (\ref{420}). We discussed already one possibility for the 
toroidal case, namely to take negative charges. This means
\be485
\tilde H^1 = \tilde h^1 - {|p^1| \over r} \ . 
\ee
Therefore, for the special radius $r={|p^1|\over\tilde h^1}=
-i{|p^1|\over 2X^1_{\infty}}$  (i.e.\ on an $S_2$ sphere) the
3-cycles collapses and at special points in moduli space, where
\be486
q_0 X^0_{\infty} = - F_1^{\infty} |p^1|
\ee
the BPS state (black hole) becomes massless (note $F_1^{\infty}$ is
negative). But if this relation is not fulfilled the black
hole is still massive. This seems to be in contradiction to
Strominger's statement \cite{strominger} that a vanishing cycle is always
related to a massless black hole. The reason is that we still
kept $q_0$ as independent charge, i.e.\ we still have a
bound state of 2 objects. Setting this charge to zero,
the mass vanish iff $F_1^{\infty} \sim \tilde h^1 =0$
for arbitrary magnetic charge. Doing this ($q_0 = \tilde h^1=0$) in 
(\ref{490}), we obtain as  massless black hole solution 
\be487
e^{-2U} = 1 + {(p^1)^2 \over 8 \pi r^2} \log {c (p^1)^2 \over r^2 } \pm ..
\quad , \quad z^1 = -i {\sqrt {c} \, p^1 \over r ( 1 \mp ..)} 
\quad , \quad  F_{mn} = \epsilon_{mnp} \partial_p \, {p^1 \over r} \ .
\ee
where we used $h_0 = 2 \sqrt{c}$,  in order to have an asymptotic
Minkowski space and $\pm\cdots $ indicate higher powers in $1/r$.
The exact solution is given by (\ref{490}), if one inserts the
exact solution for $\lambda$ from eq.\ (\ref{470}).
This is dual to the electric  configuration discussed by
Strominger. In order to keep the sign of $z^1$, the charge $p^1$
has to be positive. 

To our knowledge, this solution has not been discussed before. 
Therefore, let us add some
further comments. The main property of this massless solution is that it
carries only one charge and has a shrinking internal
3-cycle at spatial infinity ($z^1 \to 0$ for $r \to \infty$). We have
expanded the solution around this vanishing cycle, but what about the
black hole singularity/horizon? To discuss this question we have to
approach the point $r=0$. Since $X^1$ blows up in this limit, our
expansion breaks down and we have to take a different expansion in order
to find the new parameter $\lambda$. The main contribution in this region
will come from the cubic intersection part and as solution near $r=0$ one
obtains $\lambda \sim \sqrt{(\tilde H^1)^3 / H_0}$. Keeping in mind that
$H_0 = 2 \sqrt{c}$ for this solution, we find $\exp\{ -2U\} \sim (\tilde
H^1)^{3/2}$, i.e.\ this solution has a singular horizon (vanishing area).
This however, is a typical example of a compactification singularity. In
the mirror-mapped IIA solution, this black hole appears as
compactification of the 5-d magnetic string, which is a wrapped
$M5$-brane. Turning of the electric charges (see before eq.\
(\ref{487})), means that there are no momentum modes traveling along this
string and after compactification it becomes a singular black hole
(see also the next footnote).


\bigskip

{\bf \large 6. Summary and discussion}

The aim of this paper was to discuss the strong relationship between the
4-dimensional space time and the internal space. The general picture is
shown in figure 1: for general extremal $N=2$ black hole solutions the
moduli, i.e. the cycles of the internal Calabi-Yau space vary over
space-time. We found two solutions, describing Calabi-Yau phase
transitions. In the first solution, at special radii in (uncompactified)
space the Calabi-Yau degenerates and topology change can occur. These
special radii can be seen as the positions of the massless extremal black
holes. This solution carries two charges, where one of them has to be
negative. The second solution is described by only one positive charge and
the phase transition point is at spatial infinity. This solution
represents a (so far unknown) massless black hole.

\begin{figure}[t]
\hfill \begin{minipage}{15cm}
\includegraphics[width=60mm,angle=-90]{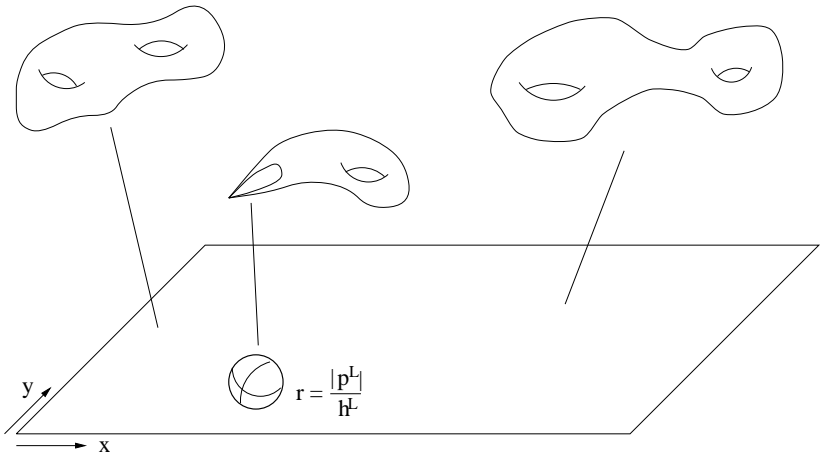}
\end{minipage}
\caption{This figure shows
the Calabi-Yau space varying over the
three-dimensional space.
In type IIB strings,  at the conifold point some 3-cycles of the Calabi-Yau
space vanish and are replaced by topologically different cycles.
We argue, that this happens on $S_2$ spheres in space time where
harmonic functions vanish. These spheres can be seen as
domain walls that separate topological different vacua.
}
\end{figure}

What can we say about the black hole singularity/horizon? For torus
compactification the space-time solution becomes singular at the internal
degeneration loci (see eq.\ (\ref{230})). In this case, any vanishing
cycle is accompanied by diverging cycles (decompactification).
%
However, for Calabi-Yau compactification, the black hole
metric stays finite at the loci of collapsing cycles.
%
One reason why the black holes are 
non-singular in Calabi-Yau compactifications is that
in contrast to the torus compactification,
the branes are wrapped in a topological non-trivial manifold, which
especially means that a brane can intersect with itself.
For the 4-d black hole these self-intersections have the same
consequence as intersection of different branes - the black hole becomes
less singular. This stabilisation due to brane-intersection especially 
applies for the flop transition in type IIA compactifications. 
As long as one keeps $H_0$
non-trivial,\footnote{The special role of $H_0$ can be understood by
looking on the IIA side. It parameterises KK modes related to the $S_1$
in $M$ theory compactification. This circle cannot be stabilised by
effects coming from the Calabi-Yau and turning off these KK-modes, the
4-d black hole becomes singular, e.g.\ the black hole
(\ref{487}).} any 10-d configuration yields a
non-singular black hole upon Calabi-Yau compactification. 
Near the horizon we always get the non-singular Reissner-Nordstr{\o}m
solution with stabilised scalars. This is true as long as we restrict
ourselves on the static case. For rotating black holes one has a ring
singularity and these arguments do not hold. This coincides with
the solutions discussed in this paper. Only the single charged
solution, describing the phase transition at infinity, is
singular at $r=0$, since $H_0$ is trivial in this case. 
The other solutions remain
finite, even at points where the internal cycle collapses. 

Another interesting question is what happens  if we go beyond
the special radius in space where the Calabi-Yau degenerates.
For the type IIA flop transition it seems that there is no problem for 
the modulus, i.e
the harmonic function
to become negative beyond the wall; namely one is just entering a new
K\"ahler cone. However, for the conifold transition, it is
not so clear how to continue our solutions beyond the special radius,
since at the conifold point the Hodge numbers change, and new harmonic
functions emerge that correspond to the new cycles.


It is even not quite clear whether one should call our
solutions black holes. Instead,
it seems attractive to see these objects (spheres) as domain walls that
separate two (topologically) different vacua of the theory. Since they
are massless, they do not contribute to the total ADM mass.  For the
torus compactification these spheres are singular ($R^2_{curv} \sim (r-
|p|)^{-8}$ for two negative charges), however Calabi-Yau corrections
smooth out the singularity (see (\ref{490})). So, if one would know the
solution from ``the inside'' it should be possible to 
connect the two regions. 
In this sense the massless states are a ``door'' to another world.


\bigskip

{\bf Acknowledgments}

We thank G. Curio and T. Mohaupt for useful conversations.
K.B. thanks John Schwarz and the group at Caltech for useful
discussions. In addition, he would like to thank the Stanford
University for its hospitality. 
The work is supported by 
the Deutsche Forschungsgemeinschaft (DFG) and by the European Commision TMR
programme ERBFMRX-CT96-0045.  W. A. S is partially supported by
DESY-Zeuthen.



\begin{thebibliography}{99}

\bibitem{aspgremor}
P. Aspinwall, B. Greene and D. Morrison, Nucl. Phys. {\bf B416} (1994) 414;
{\tt hep-th/9309097}.

\bibitem{witten2}
E. Witten, \NP471 (1996) 195, Nucl. Phys. {\bf B471} (1996) 195;
{\tt hep-th/9603150}.




\bibitem{witten} E. Witten, Nucl. Phys. {\bf B403} (1993) 159,
{\tt hep-th/9301042}.

\bibitem{conifolds} C.H. Clemens, Adv. Math. {\bf 47} (1983) 107;
R. Friedman, Math. Ann. {\bf 274} (1986) 671;\\
F. Hirzebruch, ``Some Examples of Threefolds with Trivial Canonical Bundle",
in Gesammelte Abhandlungen, Band II, Springer-Verlag (1987) pp 757;\\
P. Candelas and X. de la Ossa, Nucl. Phys. {\bf B342} (1990) 246;\\
P. Candelas, P. Green and T. H\"ubsch, Nucl. Phys. {\bf B330} (1990) 49;\\
P. Candelas, X. de la Ossa, P. Green and L. Parkes, Nucl. Phys. {\bf B359}
(1991) 21.

\bibitem{strominger}
A. Strominger, Nucl. Phys. {\bf B451} (1995) 96, {\tt hep-th/9504090}.

\bibitem{gremorstro}
B. Greene, D. Morrison and A. Strominger, Nucl. Phys. {\bf B451} (1995) 109,
{\tt hep-th/9504145}.


\bibitem{sw}
N. Seiberg and E. Witten, Nucl. Phys. {\bf B426} (1994) 19,
{\tt hep-th/9407087}.

\bibitem{be/ca}
S. Ferrara, R. Kallosh and A. Strominger, 
{\it Phys. Rev.} {\bf D52} 
(1995) 5412, { hep-th/9508072};\\
S. Ferrara and R. Kallosh,
{\it Phys. Rev.} {\bf D54} (1996) 1514, {\tt hep-th/9602136};\\
S. Ferrara, R. Kallosh
{\it Phys.Rev.} {\bf D54} (1996) 1525, {\tt hep-th/9603090};\\
R. Kallosh, M. Shmakova and W.K. Wong, 
{\it Phys. Rev.} {\bf D54}(1996) 6284, 
{ hep-th/9607077};\\
S. Ferrara, G.W. Gibbons, R. Kallosh, {\it
Black holes and critical points in moduli space}, {\tt hep-th/9702103};\\
K. Behrndt, R. Kallosh, J. Rahmfeld, M. Shmakova and 
W.K. Wong,
{\it Phys. Rev.} {\bf D54} (1996) 6293, { hep-th/9608059};\\
G. Lopes Cardoso, D. L\"ust and T. Mohaupt, {\it Phys. Lett.} {\bf B388}
(1996) 266, {\tt hep-th/9608099};\\
K. Behrndt, G. Lopes Cardoso, B. de Wit, R. Kallosh, D. L\"ust and 
T. Mohaupt, \NP488 (1997) 236, {\tt hep-th/9610105}.

\bibitem{be/mo}
K. Behrndt, T. Mohaupt, {\em Entropy of N=2 black holes and 
their M-brane description}, {\tt hep-th/9611140}; \\ 
K. Behrndt, G. Lopes Cardoso, I. Gaida, {\em Quantum N=2 supersymmetric 
black holes in the S-T model}, {\tt ep-th/9704095}.

\bibitem{sa}
W. A. Sabra, {General static $N=2$ black holes}, {\tt hep-th/9703101};\hfill
\break W. A. Sabra, {\it Black holes in $N=2$ supergravity and harmonic functions}, 
{\tt hep-th/9704147}.

\bibitem{be/lu}
K.\ Behrndt, D.\ L\"ust and W.A.\ Sabra, {\em Stationary solutions of 
N=2 supergravity}, {\tt hep-th/9705169}.

\bibitem{hyper}
K. Behrndt, I. Gaida, D. L\"ust, S. Mahapatra and T. Mohaupt,
{\em From Type IIA Black Holes to T-dual Type IIB D-Instantons in $N=2$, $D=4$
Supergravity}, {\tt hep-th/9706096}.


\bibitem{aft}
I. Antoniadis, S. Ferrara and T.R. Taylor, Nucl. Phys.
{\bf B460} (1996) 489, {\tt hep-th/9511108}.

\bibitem{louis}
J. Louis, K. Sonnenschein, S. Theisen and S. Yankielowicz, Nucl. Phys.
{\bf B480} (1996) 185; {\tt hep-th/9606049}.

\bibitem{fkm}
S. Ferrara, R.R. Khuri and R. Minasian, Phys. Lett. {\bf B375} (1996) 81;
{\tt hep-th/9602102}.

\bibitem{chou}
A. Chou, R. Kallosh, J. Rahmfeld, S.-J.Rey, M. Shmakova and W. Kai Wong,
{\em 
Critical points and phase transitions in 5-D compactifications of M theory},
{\ttfamily hep-th/9704142}





\bibitem{an/fe}
I. Antoniadis, S. Ferrara, R. Minasian and K.S. Narain,
{\em $R^4$ Couplings in $M$ and type II theories on Calabi Yau
spaces}, {\tt hep-th/9707013}.






\end{thebibliography}
\end{document}